\documentclass[useAMS,usenatbib,onecolumn]{mn2e}
\usepackage{graphicx}
\usepackage{multirow}
\usepackage{datetime}
\usepackage{array} 
\title[Small carbon chains in circumstellar envelopes]{Small carbon chains in circumstellar envelopes}

\author[R. J. Hargreaves, K. Hinkle and P. F. Bernath]{R. J. Hargreaves$^{1}$\thanks{E-mail: rhargrea@odu.edu}, K. Hinkle$^{2}$ and P. F. Bernath$^{1,}$$^{3}$\\
$^{1}$Department of Chemistry, Old Dominion University, 4541 Hampton Boulevard, Norfolk, VA 23529, USA\\
$^{2}$NOAO, 950 North Cherry Avenue, Tucson, AZ 85719, USA\\
$^{3}$Department of Chemistry, University of York, Heslington, York, YO10 5DD, UK}

\begin{document}

\date{Accepted 19 August, 2014. Received 15 August, 2014; in original form 10 July, 2014.}

\pagerange{\pageref{firstpage}--\pageref{lastpage}} \pubyear{2014}

\maketitle

\label{firstpage}

\begin{abstract}
Observations were made for a number of carbon-rich circumstellar envelopes using the Phoenix spectrograph on the Gemini South telescope to determine the abundance of small carbon chain molecules. Vibration-rotation lines of the $\nu_{3}$ antisymmetric stretch of C$_{3}$ near 2040 cm$^{-1}$ (4.902 $\mu$m) have been used to determine the column density for four carbon-rich circumstellar envelopes: CRL 865, CRL 1922, CRL 2023 and IRC +10216. We additionally calculate the column density of C$_{5}$ for IRC +10216, and provide an upper limit for 5 more objects. An upper limit estimate for the C$_{7}$ column density is also provided for IRC+10216. A comparison of these column densities suggest a revision to current circumstellar chemical models may be needed.
\end{abstract}


\section{Introduction}

Obscured carbon stars provide us with fascinating environments to study the spectacular mass loss stage in the evolution of a typical star. In the standard model, stellar pulsation coupled with circumstellar wind ejects mass from the stellar surface into the interstellar medium. The primary consequence of stellar mass loss is the formation of a circumstellar envelope and during this process the ejected material undergoes several stages of chemical and physical processing \citep{2008PhST..133a4028O}. The circumstellar material is made up of molecular gas and `dusty' grains that can provide surfaces for additional processing and grain growth. As the circumstellar shell expands, photochemical processes also become important in the breakdown and formation of further molecules \citep{2006PNAS..10312274Z}.

It is well established that the carbon to oxygen (C/O) ratio dictates the molecular species in circumstellar shells due to the high stability of the CO molecule \citep{1973AANDA....23..411T}. This allows a broad variety of molecular species to form from the remaining oxygen-rich or carbon-rich material in circumstellar shells of asymptotic giant branch (AGB) stars \citep{2008PhST..133a4028O}. IRC +10216 (CRL 1381) is carbon-rich and is the best studied example of a circumstellar shell \citep{1996ARAANDA..34..241G, 2010AANDA...521L...8C, 2012AANDA...543A..73M} with more than 80 molecules detected so far, typically from pure rotational spectra \citep{2012AANDA...543A..48A}. Approximately 80\% of these molecules are carbon containing species \citep{2006PNAS..10312274Z}, although an interesting oxygen chemistry has also been shown to exist \citep{2006ApJ...650..374A}, thus making circumstellar shells a challenging environment for physical and chemical models \citep{2012AANDA...545A..12C}.

The smallest pure carbon chains (C$_{2}$ and C$_{3}$) are ubiquitous throughout the interstellar medium \citep{2003ApJ...595..235A} and have been detected in diffuse interstellar clouds \citep{1977ApJ...216L..49S, 2000ApJ...534L.199C}, comets \citep{2000ARAANDA..38..427E} and star forming cores \citep{2012AANDA...546A..75M}. Pure carbon chains do not possess a permanent dipole moment and therefore have no allowed pure rotational transitions.  A thick `dusty' circumstellar shroud means there is little flux in the visible region and absorptions from electronic transitions cannot be observed due to scattering. However, this makes circumstellar envelopes ideal for observing vibration-rotation bands or low-lying electronic transitions in the infrared. The C$_{2}$ \citep{1975ApJ...198L.135C}, C$_{3}$ \citep{1988Sci...241.1319H} and C$_{5}$ \citep{1989Sci...244..562B} pure carbon chains have been detected in IRC +10216. Longer pure carbon chains have yet to be identified but their derivatives, such as HC$_{n}$ ($n=1-8$), H$_{2}$C$_{n}$ ($n=2-4$) and HC$_{2n}$N ($n=1-5$,) are prominent species in circumstellar shells and interstellar clouds \citep{2000ARAANDA..38..427E}. Small carbon chains (C$_{n}$) are therefore believed to be the fundamental building blocks of larger carbon containing species, such as polycyclic aromatic hydrocarbons (PAHs), cyanopolyynes and fullerenes \citep{2000ARAANDA..38..427E, 10.1021/cr970004v}. Indeed, the pure carbon molecules, C$_{60}$ and C$_{70}$, have been observed in planetary nebulae \citep{2010Sci...329.1180C}. Small carbon chains are also important in flame chemistry \citep{1977JChPh..66.3300B} and soot formation \citep{2010MolPh.108.1013G}.

Understanding circumstellar chemistry relies heavily on the archetypical object IRC +10216 due to its proximity and large mass loss \citep{2008PhST..133a4028O,2012AANDA...543A..48A}. Spectral surveys have been preformed on circumstellar shells around other carbon stars, such CIT 6 \citep{2009ApJ...691.1660Z} and CRL 3068 \citep{2009ApJ...700.1262Z}, which both show a similar chemical composition to IRC +10216. However, as a carbon-rich AGB star evolves, the chemical composition changes primarily due to photochemical processes \citep{2011IAUS..280..237C}. The circumstellar winds transport molecules and dust into the envelope where chemical pathways are thought to lead to the formation of carbon chains and aromatic rings \citep{2006AANDA...456.1001C, 2012AANDA...545A..12C}. These molecules are then broken down as a part of the AGB shell evolution towards planetary nebulae. Determining the abundance of small chains in circumstellar shells allow chemical models to be refined, whilst comparing carbon chain abundances between carbon-rich circumstellar shells determines the appropriateness of IRC +10216 as the prototypical example.

Vibration-rotation transitions of pure carbon chains allow the detection of these molecules in circumstellar envelopes when high-resolution experimental data is available. The $\nu_{3}$ antisymmetric stretch of C$_{3}$ has been the focus of a number of experimental and theoretical studies \citep{1988JChPh..89.3491M, 1993JChPh..98.7757M, 1997JMoSp.183..129S}, and has a band center at 2040.019 cm$^{-1}$ \citep[vacuum wavelength\footnote{All wavelegths are given for a vacuum.} 4.901915 $\mu$m;][]{2013JPCA..117.3332K}. The $\nu_{3}$ mode of C$_{5}$ has also been investigated \citep{1989JChPh..90..595V, 1989JChPh..91.2140M, 2007JChPh.127o4318M} with a band center at 2169.442 cm$^{-1}$ \citep[4.609480 $\mu$m;][]{1989Sci...244..562B}. C$_{7}$ has yet to be detected in circumstellar envelopes, but the strongest vibrational band, the $\nu_{4}$ antisymmetric stretch at 2138.314 cm$^{-1}$ \citep[4.676582 $\mu$m;][]{2007JChPh.127a4313N}, has been measured in the laboratory \citep{1991JChPh..94.1724H, 2013JChPh.139f4301M}. The close proximity of these fundamental frequencies (within 0.3 $\mu$m) allow all three molecules to be studied in circumstellar shells using the same high-resolution instrument.

\section{Objects and Observations}

High-resolution infrared spectra were obtained of target carbon stars with known circumstellar shells using the 8.1\,m Gemini South telescope and the NOAO Phoenix spectrometer \citep{1998SPIE.3354..810H}. The observations (GS-2010A-Q-74) were carried out during the first half of 2010 and the program stars are listed in Table~\ref{tab-observations}. All program stars are tip AGB Mira variables and their K band flux varies by approximately one magnitude. In addition to the program stars, a number of hotter reference stars were also observed for the removal of telluric features.

\begin{table*}
  \centering
  \caption{Observations using the Phoenix Spectrometer on Gemini South (GS-2010A-Q-74)}
  \label{tab-observations}
  \centering
  \begin{tabular}{@{}lccccccccc@{}}
  \hline
  \rule{0pt}{1ex}  \\
            &      &  \multicolumn{3}{c}{Magnitude}  &    \multicolumn{3}{c}{Observation dates in 2010}     & Magnitude \\
Object      & IRAS &           &          &          &    \multicolumn{3}{c}{(integration time in minutes)} & References\\
            &      &       $J$ &      $H$ &      $K$ &  C$_{3}$ Region &  C$_{5}$ Region &  C$_{7}$ Region  &           \\
  \rule{0pt}{1ex}  \\
 \hline
  \rule{0pt}{1ex}  \\
 \textit{Program Stars}:\\
  \rule{0pt}{1ex}  \\
 CRL 865    & $06012+0726$ &  $-$ & 11.3 &  7.7 &  24 Feb (24) &  21 Feb (16)   &  23 Feb (32)    & 1 \\
 IRC +10216 & $09452+1330$ &  7.3 &  4.0 &  1.2 &  03 Mar (0.8)&  25 Feb (5.3)  &  23 Feb (8)     & 1 \\
 CRL 1922   & $17049-2440$ & 12.2 &  9.2 &  6.3 &  12 Jun (8)  &  01 Mar (8)    &  26 Apr (16)    & 2 \\
 CRL 2023   & $17512-2548$ & 13.2 &  9.5 &  6.5 &  12 Jun (16) &  04 Mar (38)   &  22 May (32)    & 2 \\
 CRL 2178   & $18288-0837$ & 11.7 &  8.0 &  5.2 &       $-$    &  23 May (52)   &  27 Jun (21.3)  & 2 \\
 CRL 3099   & $23257+1038$ &  $-$ & 10.3 &  7.1 &       $-$    &  27 Jun (24)   &       $-$       & 1 \\
  \rule{0pt}{1ex}  \\
 \hline
  \rule{0pt}{1ex}  \\
 \textit{Reference Stars}:\\
  \rule{0pt}{1ex}  \\
 HIP 22509   & $04478+0848$ &  4.26 &  4.21 &  4.17 &        $-$   &  21 Feb (16)  &       $-$     & 3, 4 \\
 HR 1790     & $05224+0618$ &  2.17 &  2.24 &  2.32 &  24 Feb (38) &       $-$     &       $-$     &    3 \\
 HR 2491     & $06429-1639$ & -1.36 & -1.33 & -1.35 &        $-$   &  25 Feb (2.7) &  23 Feb (1.3) &    3 \\
 HR 4662     & $12132-1715$ &  2.76 &  2.83 &  2.81 &  12 Jun (50) &       $-$     &       $-$     &    3 \\
 HR 5267     & $14002-6008$ &  1.17 &  1.21 &  1.28 &  03 Mar (96) &  01 Mar (48)  &  27 Jun (48)  &    4 \\
  \rule{0pt}{1ex}  \\
 \multirow{3}{*}{HR 6553} & \multirow{3}{*}{$17337-4258$} & \multirow{3}{*}{1.07} & \multirow{3}{*}{0.87} & \multirow{3}{*}{0.84} & \multirow{3}{*}{$-$} &  04 Mar (24)   &  26 Apr (40) & \multirow{3}{*}{4}   \\
                          &                               &                       &                       &                       &                      &  23 May (40)   &  22 May (40) &                           \\
                          &                               &                       &                       &                       &                      &  27 Jun (53.3) &              &                           \\
  \rule{0pt}{1ex}  \\
 \hline
  \rule{0pt}{1ex}  \\
\multicolumn{8}{l}{\textit{References.} (1) \citet{2006MNRAS.369..751W}; (2) \citet{2012AJ....143...36C}; (3) \citet{2002yCat.2237....0D}; (4) \citet{2003yCat.2246....0C}.}\\
\end{tabular}
\end{table*}

The Phoenix spectrometer is a cryogenically-cooled echelle spectrograph that uses order-separating filters to isolate individual echelle orders. The detector is a $1024\times1024$ InSb Aladdin II array. A 2 pixel slit width was used resulting in a spectral resolving power of $R=\lambda/\Delta\lambda \sim 70,000$. Three spectral regions were observed centred on 2045 cm$^{-1}$ (4.890 $\mu$m), 2168 cm$^{-1}$ (4.613 $\mu$m) and 2138 cm$^{-1}$ (4.677 $\mu$m) to cover the corresponding C$_{3}$, C$_{5}$ and C$_{7}$ vibration bands. The size of the detector along the dispersion direction limits the wavelength coverage for a single observation to $\sim$0.5\% ($\sim$10.5 cm$^{-1}$ at 2100 cm$^{-1}$).

The observations and reductions employed standard thermal infrared techniques \citep{1992..Joyc..Conf} and used the IRAF software package.\footnote{IRAF software is distributed by the National Optical Astronomy Observatories under contract with the National Science Foundation.} The spectra were wavelength calibrated using telluric lines of CO from HITRAN \citep{2013JQSRT.130....4R} and a typical RMS uncertainty in the wavenumber scale is less than 0.005 cm$^{-1}$. The same HITRAN lines were used in the calculation of the telluric model atmospheres (see Section 3).

\section{Model Atmospheres and Telluric Removal}

The reference star observations, listed in Table~\ref{tab-observations}, consistently have a lower signal-to-noise compared to the object spectra for which the typical signal-to-noise ratio is $\sim$500. When the reference spectra are used to correct for atmospheric absorption, the object spectra are degraded. Recently, \citet{2010AANDA...524A..11S} have demonstrated the advantage of using synthetically produced absorption atmospheric spectra to remove telluric features from spectra taken with the CRIRES spectrograph.

The Reference Forward Model\footnote{Reference Forward Model, RFM (v4.30), A. Dudhia, University of Oxford, http://www.atm.ox.ac.uk/RFM} (RFM) is a line-by-line radiative transfer model used to calculate atmospheric transmission spectra and is often used to interpret satellite observations \citep{2008ACP.....8.2151F}. RFM can be used to compute the atmospheric transmission for an observer at any zenith angle, making RFM well suited to simulating astronomical reference spectra for telescope observations with differing airmass. The primary advantage of using RFM in place of the observed reference spectra is to eradicate noise introduced during the removal of telluric lines.

The synthetic spectra computed for this study were calculated using RFM v$4.30$, which uses the HITRAN 2012 database for molecular line parameters \citep{2013JQSRT.130....4R}. The Michelson Interferometer for Passive Atmospheric Sounding (MIPAS) night time model atmospheric profiles (\textit{ngt.atm}) were used with the CO$_{2}$ profile scaled to 2010 concentrations and the H$_{2}$O profile scaled to match observation. A Gaussian instrument lineshape (FWHM = 0.03 cm$^{-1}$) was also used to account for the lineshape of the Phoenix spectrometer. In addition, telescope temperature measurements were combined with National Centers for Environmental Prediction (NCEP) temperature profiles for Gemini South during the period of measurement to adjust the temperature profile below 25 km (i.e., the troposphere and lower stratosphere). Above 25 km, the MIPAS \textit{ngt.atm} temperature profile remained unchanged. Since the H$_{2}$O absorptions and weather conditions were different for each observation, it was necessary to produce an RFM synthetic spectrum for each program observation to replace the reference star observations.

\begin{figure}
 \centering
 \hspace*{0.5\textwidth} 
 \includegraphics[width=0.45\textwidth, angle=90]{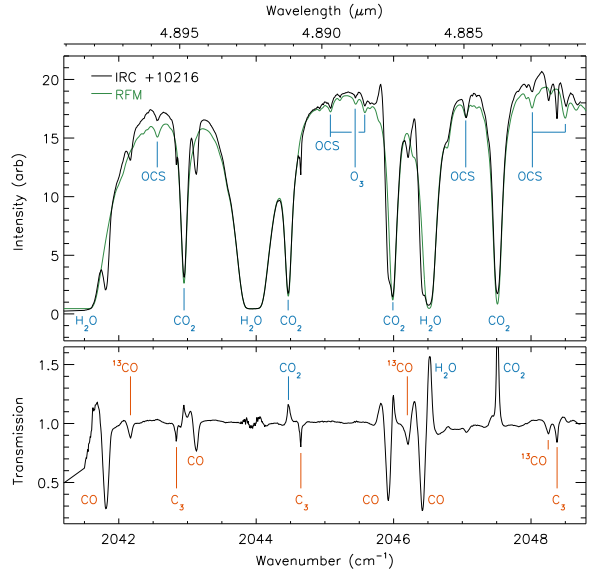}
 \caption{Calibrated spectrum of IRC +10216 with the corresponding RFM synthetic spectrum (upper panel) and resulting ratioed IRC +10216 spectrum with telluric lines removed (lower panel). Molecular absorption from telluric lines (blue) and the circumstellar shell (orange) have been indicated.}
 \label{REFvsRFM}
\end{figure}

The IRAF routine TELLURIC was used to remove telluric features and to ratio program stars with the RFM synthetic spectra. Figure~\ref{REFvsRFM} shows the observed C$_{3}$ spectral region (approximately $2041-2049$ cm$^{-1}$) for IRC +10216 and the corresponding RFM synthetic spectrum, with strongly absorbing telluric features indicated. The resulting spectrum for IRC +10216 is also shown in Figure~\ref{REFvsRFM} with circumstellar molecular lines identified. Whilst complete removal of some telluric features is not achieved (e.g., H$_{2}$O and CO$_{2}$), the result is an improvement on telluric removal when the reference star spectra are used. The C$_{3}$ absorptions are sharp and of moderate intensity ($\sim$15\% absorption) and can be identified before telluric features have been removed. The CO lines are unlike C$_{3}$ because they have characteristic P-Cygni profiles due to the distribution of CO throughout the circumstellar envelope at different radial velocities \citep{1988ApJ...326..832K}.

\section{Results}

\subsection{C$_{3}$}

Telluric removal was carried out on all observations from Table~\ref{tab-observations} using RFM synthetic spectra. The observed C$_{3}$ spectral region contained the \textit{R}(2), \textit{R}(4), \textit{R}(6) and \textit{R}(8) rotational lines of the $\nu_{3}$ antisymmetric stretch at 2042.66, 2044.48, 2046.32 and 2048.20 cm$^{-1}$, respectively. Table~\ref{tab-results} summarises the line identifications for each of the program stars. Missing $R$-branch lines are due to absorption coincidences with telluric features due to relative velocity shifts. From these observations, the column densities and local standard of rest velocity shifts were calculated.

\begin{table*}
\centering
  \caption{Summary of C$_{3}$ line identifications}
  \label{tab-results}
  \centering
  \begin{tabular}{@{}lccccccccc@{}}
  \hline
  \rule{0pt}{1ex}  \\
   Program  & \multicolumn{4}{c}{Observed Lines$^{a}$ (cm$^{-1}$)} &  Obs$-$Ref$^{b}$  &    Temperature   & Partition &     Column             \\
   Star     &            &            &            &            &   (cm$^{-1}$)  &      (K)         & Function  &     Density            \\
            &   $R$(2)   &   $R$(4)   &   $R$(6)   &   $R$(8)   &                &                  &           & (cm$^{-2}$)   \\
  \rule{0pt}{1ex}  \\
 \hline
  \rule{0pt}{1ex}  \\
 CRL 865    &  2042.188  &     $-$    &  2045.845  &  2047.718  &   $-$0.480    &    44.3           &   40.3    &  $9.6\times10^{14}$  \\
 IRC +10216 &  2042.835  &  2044.647  &     $-$    &  2048.377  &   $+$0.172    &    49.1           &   46.9    &  $8.8\times10^{14}$  \\
 CRL 1922   &  2042.888  &  2044.670  &     $-$    &  2048.423  &   $+$0.222    &    29.3           &   24.8    &  $6.8\times10^{14}$  \\
 CRL 2023   &     $-$    &  2044.714  &     $-$    &  2048.432  &   $+$0.233    &    50.0$^{c}$     &   57.4    &  $3.8\times10^{14}$  \\
  \rule{0pt}{1ex}  \\
\hline
  \rule{0pt}{1ex}  \\
\multicolumn{9}{p{14cm}}{$^{a}$ in telluric frame of reference.} \\
\multicolumn{9}{p{14cm}}{$^{b}$ average observed$-$reference \citep[references determined from][]{2013JPCA..117.3332K}.} \\
\multicolumn{9}{p{14cm}}{$^{c}$ estimated temperature.}\\
\end{tabular}
\end{table*}

The ratioed program star spectra were normalised and the baseline was flattened to account for small variations over the observed region. The C$_{3}$ absorption peaks were fit to a Gaussian lineshape using OriginPro 8.0. The rotational temperature of C$_{3}$ within the circumstellar envelope can be calculated from the relationship between the absorption intensity $I$, to temperature $T$ \citep{1989..Herz..book}, since
\begin{equation}
   I \propto (J' + J'' + 1)e^{-B'' (J'' + 1) hc / kT },
\end{equation}
where $B'' = 0.436$ cm$^{-1}$ \citep{2013JPCA..117.3332K}. The slope obtained from $\ln[I/(J' + J'' + 1)]$ versus $J''(J'' + 1)$ yields $-{B'' hc}/{kT}$ and thus the temperature. The C$_{3}$ temperature for each program star is given in Table~\ref{tab-results} based on the identification of each C$_{3}$ peak. For CRL 2023, only two lines were observed (giving a rotational temperature of 133 K) therefore an estimated temperature of 50 K was used instead.

The molecular C$_{3}$ column densities can be obtained from the Beer-Lambert law
\begin{equation}
   I = I_{0} e^{-S' g(\nu-\nu_{10})Nl},
\end{equation}
where $S'$ is the `line strength', $I$ is the line intensity and $g(\nu-\nu_{10})$ is the line shape function \citep{2005..Bern..book}. Integrating over the whole line and rearranging for column density $Nl$ gives
\begin{equation}
   Nl = \frac{1}{S'} \int \ln(I_{0}/I) d\nu,
\end{equation}
where the integral is given by the area of each C$_{3}$ absorption peak and line strength, $S'$, is obtained from
\begin{equation}
   S' = \frac{2 \pi^{2} \nu_{10} S_{J' J''}}{3 \epsilon_{0} h c Q} e^{-E_{0}/kT} (1 - e^{h \nu / kT}).
\end{equation}

The partition functions $Q$, was calculated individually for each program star (Table~\ref{tab-results}) based on the temperature and the contribution due to the rotational and vibrational components. The rotational line strength, $S_{J' J''}$, was calculated from the transition dipole moment $M_{\nu' \nu''}$ \citep{2005..Bern..book} as
\begin{equation}
   S_{J' J''} = |M_{\nu' \nu''}|^{2} \cdot \textrm{HLF},
\end{equation}
where HLF is the H\"{o}nl-London Factor (HLF = $J'' + 1$ for $R$-branch lines of C$_{3}$) and $M_{\nu' \nu''} = 0.35$ debye \citep{1992JChPh..97.3399J}.

The spectral simulation program PGOPHER\footnote{PGOPHER (v8.0.195), a program for simulating rotational structure, C. M. Western, University of Bristol, http://pgopher.chm.bris.ac.uk/} was used to simulate the rotational spectrum of the $\nu_{3}$ band of C$_{3}$ using the calculated temperature with the constants from \citet{2013JPCA..117.3332K}. This provided the lower state energy, $E_{0}$, for each transition along with the rotational assignments.

\begin{figure}
\centering
 \includegraphics[width=0.45\textwidth]{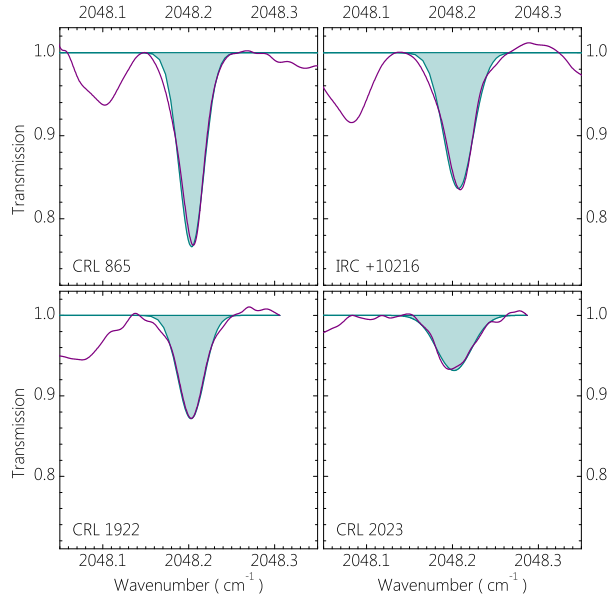}
 \caption{Calibrated identification of the $R(8)$ absorption line of the $\nu_{3}$ mode of C$_{3}$ at 2048 cm$^{-1}$ (4.882 $\mu$m) for observations in the circumstellar shells of CRL 865, IRC +10216, CRL 1922 and CRL 2023. The shaded region indicates the fitted area used to calculate the column density for each line.}
 \label{allR8}
\end{figure}

From Equation~3, the average column densities of C$_{3}$ are calculated to be $9.6\times10^{14}$, $8.8\times10^{14}$, $6.8\times10^{14}$ and $3.8\times10^{14}$ cm$^{-2}$ for CRL 865, IRC +10216, CRL 1922 and CRL 2023, respectively. These calculated column densities have an estimated error of 20\% except CRL 2023, which is about a factor of two due to the estimated temperature. Figure~\ref{allR8} displays the $R(8)$ line of the $\nu_{3}$ mode of C$_{3}$ for the program stars given in Table~\ref{tab-results}. The fitted Gaussian peak is displayed along with the area used to calculate the C$_{3}$ column density.

The sharp line shape of C$_{3}$ indicates that the molecule is contained within a narrow shell of each circumstellar envelope. Calibrating the spectra with telluric features determines the velocity shift ($v_{obs}$) of the observed line. The IRAF procedure RVCORRECT was used to calculate the velocity shifts due the motion of the Earth, Moon and Sun at the time of observation to provide the local standard of rest velocity ($v_{lsr}$). Comparing the $v_{lsr}$ values to accepted stellar velocities ($v_{obs}$) for each object, the shell expansion velocity ($v_{shell}$) can be obtained from $ v_{shell} = v_{lsr} - v_{obj} $ ; the results are summarised in Table~\ref{tab-velocities}.

\begin{table}
  \caption{Velocity components (in km s$^{-1}$) for each program star from the velocity shifted C$_{3}$ absorptions}
  \label{tab-velocities}
  \centering
  \begin{tabular}{@{}lcccc@{}}
  \hline
  \rule{0pt}{1ex}  \\
            &    Observed   &    Local Standard of  & Stellar  & Shell   \\
   Object   &    Velocity   &  Velocity  &  Rest Velocity$^{a}$  &  Velocity    \\
            &   ($v_{obs}$)   & ($v_{lsr}$) & ($v_{obj}$) & ($v_{shell}$)   \\
  \rule{0pt}{1ex}  \\
 \hline
  \rule{0pt}{1ex}  \\
 CRL 865    &     70.3      &  28.5 &  45.6 &  -17.1 \\
 IRC +10216 &    -25.2      & -41.1 & -23.2 (-26.0$^{b}$) &  -17.9 (-15.1) \\
 CRL 1922   &    -32.6      & -22.6 &  -5.1 &  -17.5 \\
 CRL 2023   &    -34.1      & -18.7 &  -    &  $\leq$ -18.7 \\
  \rule{0pt}{1ex}  \\
\hline
  \rule{0pt}{1ex}  \\
\multicolumn{5}{l}{$^{a}$ from \citet{2006MNRAS.369..783M}.} \\
\multicolumn{5}{l}{$^{b}$ from \citet{2012AANDA...543A..73M}.}
\end{tabular}
\end{table}

\subsection{C$_{5}$}

\begin{figure}
\centering
 \includegraphics[width=0.45\textwidth]{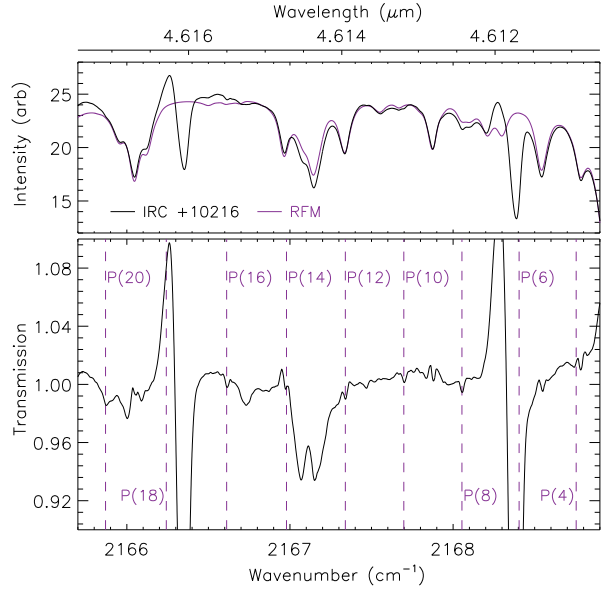}
 \caption{A section of the calibrated and velocity shifted spectrum of IRC +10216 and the corresponding RFM synthetic spectrum in the range of the $\nu_{3}$ mode of C$_{5}$ (upper panel). The ratioed IRC +10216 spectrum highlights the position of the $P$-branch absorptions shown in purple (lower panel).}
 \label{c5region}
\end{figure}

A PGOPHER simulation of the $\nu_{3}$ mode of C$_{5}$ at 50 K using the constants obtained from \citet{1989JChPh..91.2140M} and a band strength of 0.74 debye \citep{1989CPL...160..485B} predicts the $P(14)$ transition to be the most intense. Figure~\ref{c5region} displays the velocity shifted and ratioed spectrum for IRC +10216 in the proximity of the $P(14)$ transition, a number of small absorption features are coincident to the C$_{5}$ positions. There are tentative identifications of the $P(8)$, $P(10)$, $P(16)$ and $P(20)$ lines; the remaining lines are contaminated by telluric features and cannot be identified. The column density calculated from the strongest transition, $P(8)$, is $1.3\times10^{13}$ cm$^{-2}$. This calculation has been carried out for the $P(8)$ transition in the spectra of the remaining objects observed in the C$_{5}$ region. These features are not as prominent as they are in IRC +10216 and the calculated column densities are estimated upper limits as summarised in Table~\ref{tab-c5}.

\begin{table}
  \caption{Upper limit estimates for the C$_{5}$ column densities}
  \label{tab-c5}
  \centering
  \begin{tabular}{@{}lcc@{}}
  \hline
  \rule{0pt}{1ex}  \\
   \multirow{2}{*}{Object}   &     Column density upper limit$^{a}$  &  \multirow{2}{*}{Ratio$^{b}$}   \\
                             &    (cm$^{-2}$)                     &        \\
  \rule{0pt}{1ex}  \\
 \hline
  \rule{0pt}{1ex}  \\
 CRL 2023    &     $2.4\times10^{12}$   &   158   \\
 CRL 3099    &     $5.1\times10^{12}$   &   $-$   \\
 CRL 865     &     $9.4\times10^{12}$   &   102   \\
 CRL 2178    &     $1.1\times10^{13}$   &   $-$   \\
 CRL 1922    &     $1.2\times10^{13}$   &    57   \\
 IRC +10216  &     $1.3\times10^{13}$   &    68   \\
  \rule{0pt}{1ex}  \\
\hline
  \rule{0pt}{1ex}  \\
\multicolumn{3}{l}{$^{a}$ except for IRC +10216.} \\
\multicolumn{3}{l}{$^{b}$ as C$_{3}$:C$_{5}$ column density.}
\end{tabular}
\end{table}

\subsection{C$_{7}$}

\begin{figure}
\centering
 \hspace*{0.75\textwidth} 
 \includegraphics[width=0.45\textwidth, angle=90]{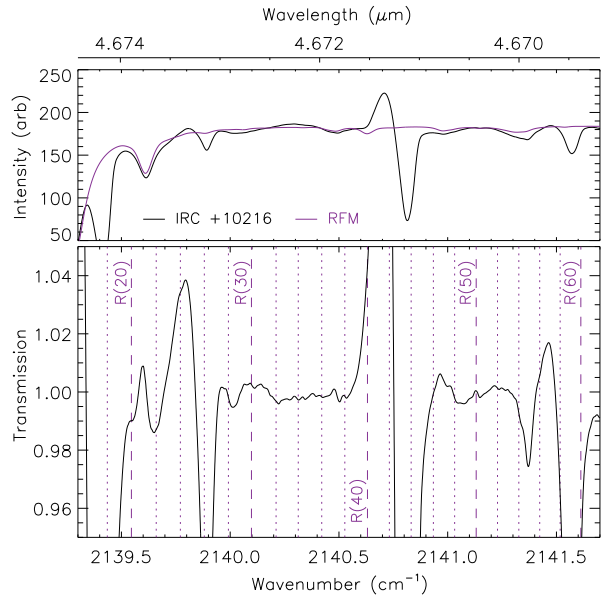}
 \caption{A section of the calibrated and velocity shifted spectrum of IRC +10216 and the corresponding RFM synthetic spectrum in the range of the $\nu_{4}$ mode of C$_{7}$ (top panel). The $R$-branch line positions are shown in purple along with the ratioed IRC +10216 spectrum (lower panel).}
 \label{c7region}
\end{figure}

A PGOPHER simulation of the $\nu_{4}$ mode of C$_{7}$ at 50 K using the constants obtained from \citet{2007JChPh.127a4313N} and a band strength of 0.72 debye determined from a band intensity of 2809 km mol$^{-1}$ \citep{1996JChPh.105.5313K} predicts the $R(24)$ transition to be the most intense. Figure~\ref{c7region} displays the reduced spectrum for IRC +10216 with predicted C$_{7}$ line positions. No assignments can be made above the noise limit. The small `peak' at 2140.12 cm$^{-1}$ near $R(30)$ has been used to estimate an upper limit of the C$_{7}$ column density, assuming the molecule is contained within a shell at 50 K. The calculated upper limit for the C$_{7}$ column density in the circumstellar envelope of IRC +10216 is $4.7\times10^{12}$ cm$^{-2}$.

\section{Discussion}

Synthetic reference spectra were necessary due to the unsatisfactory noisy reference star spectra, which in some cases had noise close to 5\%. We have demonstrated that line-by-line radiative transfer programs, such as RFM, are capable of creating suitable reference spectra for the removal of telluric features from obscured carbon stars in the infrared. These objects have complicated absorption spectra and a complete removal of telluric features is often difficult. Our synthetic spectra were created using a combination of measured and standard atmospheric conditions for the Gemini observatory. We restricted our adjustment of the standard RFM molecular atmospheric profiles to a scaling of the CO$_{2}$ and H$_{2}$O profiles. This kept the remaining molecular profiles consistent and accounts for differing humidity conditions during each observation. We also adjusted the standard temperature profile (which applies to all molecules) to coincide with the measured temperature at the telescope. For these reasons, the synthetic spectra do not completely remove telluric features, however it should be noted that this was also not possible (and worse) using the original reference star observations, most likely due to the fluctuating weather conditions during each observation run. Telescope time is a limited resource and is better spent on object acquisition, particularly if synthetic spectra are able to adequately simulate reference star observations.

The C$_{3}$ absorptions are clearly identifiable due to the strength of the $\nu_{3}$ transition. They can be assigned before telluric removal (Figure~\ref{REFvsRFM}) and were observed in the spectra of all program stars recorded in the C$_{3}$ region. These absorptions are underresolved at a resolution of 0.03 cm$^{-1}$ and the sharp transitions are broadened by the instrument line shape function; resulting in a reduced absorption depth. Nevertheless, the calculated column density of $8.8\times10^{14}$ cm$^{-2}$ for C$_{3}$ in IRC +10216 compares very well to $1.0\times10^{15}$ cm$^{-2}$ obtained by \citet{1988Sci...241.1319H}.

The same broadening occurs for our C$_{5}$ observations and a maximum absorption depth of $\sim$1\% is seen. It was only possible to calculate the column density of C$_{5}$ in IRC +10216 by using a synthetic spectrum to remove the telluric lines. C$_{5}$ has previously been observed with a column density of $9.0\times10^{13}$ cm$^{-2}$ in the same spectral region of IRC +10216 by \citet{1989Sci...244..562B} using a high-resolution Fourier transform infrared spectrometer (0.01 cm$^{-1}$ resolution); a maximum absorption depth of $\sim$3\% was observed. The calculated value of $1.3\times10^{13}$ cm$^{-2}$ is approximately 15\% of the value calculated by \citet{1989Sci...244..562B}, however \citet{1989CPL...160..485B} suggest that the \citet{1989Sci...244..562B} column density was an overestimate due to the transition dipole moment that was used. The C$_{5}$ transition dipole moment for this study was taken from \citet{1989CPL...160..485B} and the column density presented here is within 25\% of their estimated value of $5.0\times10^{13}$ cm$^{-2}$.

The C$_{5}$ absorption lines for the remaining program stars (excluding IRC +10216) are at the noise limit when synthetic reference spectra are used. Therefore, it was only possible to provide an estimate of the upper limit for C$_{5}$ in these objects. Whilst these upper limits span almost one order of magnitude, and therefore should only be used as an indication, it is interesting to note a negative correlation between the column density of C$_{5}$ and the C$_{3}$:C$_{5}$ ratio. 

\citet{1993ASPC...41..125H} have previously obtained an upper limit of $2.0\times10^{13}$ cm$^{-2}$ for a C$_{7}$ column density based on high-resolution spectra of IRC +10216. This work refines the column density upper limit for C$_{7}$ in IRC +10216 by almost one order of magnitude to $4.7\times10^{12}$ cm$^{-2}$. The column densities of C$_{3}$, C$_{5}$ and C$_{7}$ are summarised and compared to previously determined and calculated values for IRC +10216 in Table~\ref{tab-columndensities}.

\begin{figure}
\centering
 \includegraphics[width=0.45\textwidth]{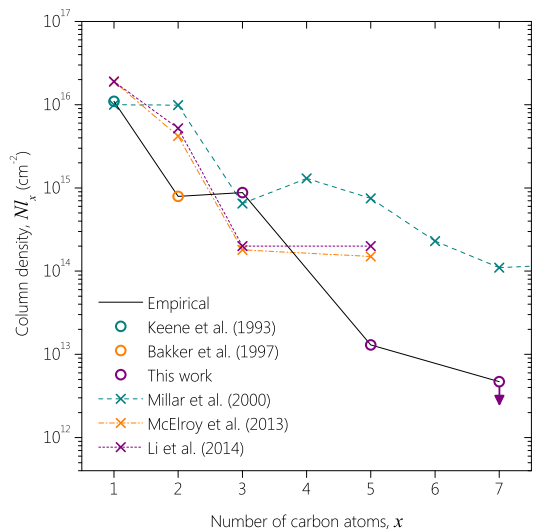}
 \caption{A comparison of the column densities of small carbon chain species in IRC +10216. Empirical column densities are shown by the solid black line and include the values for neutral carbon \citep{1993ApJ...415L.131K}, C$_{2}$ \citep{1997AANDA...323..469B} and C$_{3}$, C$_{5}$ and C$_{7}$ from this work. The dashed lines refer to calculated column densities for IRC +10216 by \citet{2000MNRAS.316..195M}, \citet{2013AANDA...550A..36M} and \citet{2014AANDA..in.press..L}.}
 \label{comparison}
\end{figure}

\begin{table}
  \caption{Summary of the observed and calculated column densities (in cm$^{-2}$) of small carbon chains in IRC +10216}
  \label{tab-columndensities}
  \centering
  \begin{tabular}{@{}lccc@{}}
  \hline
  \rule{0pt}{1ex}  \\
   Carbon   &     This  &  Observed &  Calculated \\
   chain    &     work  &  (Ref.)  &    (Ref.)  \\
  \rule{0pt}{1ex}  \\
 \hline
  \rule{0pt}{1ex}  \\
 C$_{3}$    &     $8.8\times10^{14}$      &   $1.0\times10^{15}$    (1) &   $2.0\times10^{14}$  (4)  \\
 C$_{5}$    &     $1.3\times10^{13}$      &   $9.0\times10^{13}$    (2) &   $2.0\times10^{14}$  (4) \\
 C$_{7}$    &     $\le4.7\times10^{12}$   &   $\le2.0\times10^{13}$ (3) &   $1.1\times10^{14}$  (5) \\
  \rule{0pt}{1ex}  \\
\hline
  \rule{0pt}{1ex}  \\
\multicolumn{4}{p{8cm}}{\textit{References.} (1) \citet{1988Sci...241.1319H}; (2) \citet{1989Sci...244..562B}; (3) \citet{1993ASPC...41..125H}; (4) \citet{2014AANDA..in.press..L}; (5) \citet{2000MNRAS.316..195M}. }
\end{tabular}
\end{table}

The total empirical column density in IRC +10216 for neutral carbon is $1.1\times10^{16}$ cm$^{-2}$ \citep{1993ApJ...415L.131K} and for C$_{2}$ is $7.9\times10^{14}$ cm$^{-2}$ \citep{1997AANDA...323..469B}. Figure~\ref{comparison} includes these values with those for C$_{3}$, C$_{5}$ and C$_{7}$ from this study and compares the empirical column densities to the number of carbon atoms. Also included in Figure~\ref{comparison} are calculated column densities for all pure carbon species from \citet{2000MNRAS.316..195M} and the appropriate values from \citet{2013AANDA...550A..36M}. Carbon chain column densities from a state-of-the-art model by \citet{2014AANDA..in.press..L}, that improves the N$_{2}$ and CO photodissociation rates and shield functions of \citet{2013AANDA...550A..36M}, is also shown.

The primary carbon chains growth pathways \citep{2000MNRAS.316..195M} are given by
\begin{description}
\item C$_{2}$ $+$ C$_{n}$H $\rightarrow$ C$_{n+2}$ $+$ H
\end{description}
and
\begin{description}
\item C$_{n}^{-}$ $+$ C$_{m}$ $\rightarrow$ C$_{n+m}^{-}$ $+$ $h\nu$ leading to C$_{n+m}^{-}$ $+$ $h\nu$ $\rightarrow$ C$_{n+m}$ $+$ $e^{-}$.
\end{description}
The latest models  have reduced the column densities of C$_{3}$ and C$_{5}$, however there is still a disagreement between observation and calculation. Most notably, C$_{3}$ is underestimated whereas C$_{5}$ is overestimated. The calculated values from \citet{2000MNRAS.316..195M}, \citet{2013AANDA...550A..36M} and \citet{2014AANDA..in.press..L}, shown in Figure~\ref{comparison},  predict the column densities of C$_{3}$ and C$_{5}$ to be almost equal. Our work suggests that C$_{3}$ is almost 2 orders of magnitude greater than C$_{5}$ (see Table~\ref{tab-columndensities}), which indicates a similar trend to that seen in \citet{1993ApJ...419L..41C} where the abundance of C$_{3}$ was $\sim$10 times greater than C$_{5}$. The empirical column densities for small carbon chains (particularly C$_{2}$, C$_{3}$ and C$_{5}$) indicate that the reaction rates for small carbon chains in current models are incorrect; the models are missing important reactions; or the choice of initial circumstellar conditions is not suitable.

Carbon containing molecules are ubiquitous throughout the interstellar medium and are not confined to circumstellar envelopes \citep{2009ARAANDA..47..427H}. Linear carbon chain growth is thought to eventually lead to larger species like PAHs, cyanopolyynes, fullerenes and even amino acids. Benzene (C$_{6}$H$_{6}$) has been detected in proto-planetary nebula \citep{2001ApJ...546L.123C}, cyanopolyyne species containing a chain of as many as 11 carbon atoms (HC$_{11}$N) have been observed in dark interstellar clouds \citep{1997ApJ...483L..61B} and large fullerenes, such as C$_{60}$ and C$_{70}$, have been observed in planetary nebula \citep{2010Sci...329.1180C}. \citet{2014MNRAS.437..930L} consider a revised chemical model for dark interstellar clouds that includes a number of reactions that break down larger linear carbon chains into smaller chains via reaction pathways such as
\begin{description}
\item C $+$ C$_{n}$ $\rightarrow$ C$_{3}$ $+$ C$_{n-2}$,
\end{description}
where $n \le 4$. They note that the consequence for linear carbon chains is to limit the abundance of longer chains (C$_{n>3}$) due to faster reaction rates, resulting in an accumulation of C$_{3}$, which has an estimated low reactivity. This would have the consequence of increasing the C$_{3}$ column density at the expense of C$_{5}$ (and longer chains), which would appear to improve current model calculations (Figure~\ref{comparison}).

The C$_{3}$ column densities can be used to determine an average column density of $7.3\times10^{14}$ cm$^{-2}$ for the circumstellar envelopes of carbon-rich AGB stars. However, not all carbon stars have equivalent C/O ratios or mass loss rates, which are essential in understanding the circumstellar chemistry. The C/O ratio for IRC +10216 is given as 1.4 and the mass loss rate has been calculated as $3.3\times10^{-5}$$M_{\odot}$ year$^{-1}$ \citep{2005AANDA...429..235B}, but these parameters are unavailable for the remaining objects so no correlations can be investigated. Since C$_{3}$ was observed in all objects, we believe IRC +10216 remains a prototypical example.

\citet{1997AANDA...323..469B} identified a correlation between C$_{2}$ column densities and circumstellar expansion velocity. Our results show no correlation between C$_{3}$ or C$_{5}$ column densities with the shell velocity; however if the more recent stellar velocity of -26.0 km s$^{-1}$ \citep{2012AANDA...543A..73M} is used for IRC +10216, a slight increase in column density is seen for decreasing shell velocity. Our velocities are consistent and indicate a typical circumstellar shell velocity to be approximately 17.5 km s$^{-1}$.

\section{Conclusions}

The small carbon chain C$_{3}$ has been detected in the circumstellar envelopes of CRL 865, IRC +10216, CRL 1922 and CRL 2023 by the identification of a number of vibration-rotation lines from the $\nu_{3}$ antisymmetric stretch of C$_{3}$ near 2040 cm$^{-1}$. The calculated column densities are $9.6\times10^{14}$, $8.8\times10^{14}$, $6.8\times10^{14}$ and $3.8\times10^{14}$ cm$^{-2}$, respectively. We used synthetic spectra to remove telluric features from our observations, highlighting how a line-by-line radiative transfer model can be used to remove telluric features if reference star spectra are of poor quality or unavailable. C$_{5}$ was observed in IRC +10216 with a column density of $1.3\times10^{13}$ cm$^{-2}$, and upper limits have been provided for 5 for other sources. An upper limit for the column density of C$_{7}$ in IRC +10216 was estimated to be $4.7\times10^{12}$ cm$^{-2}$. Our measurements suggest a revision to circumstellar shell models may be needed to account for small carbon chain abundances.

\section*{Acknowledgments}

This work is based on observations obtained at the Gemini Observatory, which is operated by the Association of Universities for Research in Astronomy, Inc., under a cooperative agreement with the NSF on behalf of the Gemini partnership: the National Science Foundation (United States), the National Research Council (Canada), CONICYT (Chile), the Australian Research Council (Australia), Minist\'{e}rio da Ci\^{e}ncia, Tecnologia e Inova\c{c}\~{a}o (Brazil) and Ministerio de Ciencia, Tecnolog\'{i}a e Innovaci\'{o}n Productiva (Argentina). Funding has been provided by a University of York studentship and Old Dominion University research fellowship, additional support was provided by the NASA laboratory astrophysics program. The authors would like to thank the reviewer's for their comments during the review process and Daniel Frohman for identifying appropriate band strengths to use in calculation of the C$_{5}$ and C$_{7}$ column densities.

\bsp

\label{lastpage}

\end{document}